\input psfig.sty
\documentstyle{l-aa}     

\def\mum {\mu{\rm m}}

\begin{document}
 
   \thesaurus{11.01.2; 11.07.1; 11.19.1; 13.25.2 
             }

   \title{
ROSAT All-Sky Survey observations of IRAS galaxies;\\
I. Soft X-ray and far-infrared properties 
}
 
   \author{Th. Boller$^1$, F. Bertoldi$^1$, 
           M. Dennefeld$^2$, W. Voges$^1$
          } 
   \offprints{Th. Boller
             }
 
   \institute{$^1$ Max-Planck-Institut f\"ur Extraterrestrische Physik,
                   85748 Garching, Germany\\
              $^2$ Institute d'Astrophysique (IAP), 98 bis Boulevard Arago, 
                   F-75014 Paris, France
              }
   \date{Received 9 June 1997 ; accepted 16 September 1997 }
 
   \maketitle
   \markboth{ROSAT RASS II observations of IRAS galaxies}{Th. Boller et al.} 
 
   \begin{abstract}
 
The  120,000 X-ray sources detected in the RASS~II processing 
of the ROSAT All-Sky Survey are correlated with the 
14,315 IRAS galaxies selected from the IRAS Point Source Catalogue:
372 IRAS galaxies show X-ray emission within a distance of
100 arcsec from the infrared position. By inspecting the 
structure of the X-ray emission in overlays on optical images
we quantify the likelihood that the X-rays originate from
the IRAS galaxy. 
For 197 objects the soft X-ray emission is very likely associated
with the IRAS galaxy. Their soft X-ray properties are determined
and compared with their far-infrared emission.
X-ray contour plots overlaid on Palomar Digitized Sky Survey images 
are given for each of the 372 potential identifications.
All images and tables displayed here are also available in electronic form.
 
 
\end{abstract}
 
\section{Introduction}
 
In a previous paper (Boller et al.~1992a) we presented results
of a first correlation between the ROSAT All-Sky Survey and the
14,708 extragalactic sources selected from the IRAS Point Source Catalogue
(hereafter IRAS PSC). 
This initial correlation used the first processing of
the All-Sky Survey data  by the Standard Analysis Software System (RASS~I)
(Voges et al. 1996a; Downes et al. 1994)
and resulted in a sample of 244 objects. From the distribution of
source separations (see Fig.~4 of Boller et al.~1992a) the number fraction 
of spurious sources herein was estimated to about 16$\%$.
 
Recently, a second processing of the ROSAT All-Sky Survey 
(RASS~II) 
was performed that yielded about 120,000 sources with detection 
likelihoods larger 8 (cf. Cruddace et al. (1988) for the definition 
of the detection likelihood).
The major differences of the second processing as compared with the first 
are:  
(i) the photons were not collected in strips but 
merged in 1378 sky-fields of 6.4$\times$6.4 degree, which takes full
advantage of the increasing exposure towards the ecliptic poles;
(ii) neighboring fields overlap by at least 0.23 degree in order to ensure 
the detection of sources at the field boundaries, 
which posed a problem in the first processing;
(iii) a new aspect solution
reduces the number of sources with erroneous positions and morphology.
 
In this paper we present the results of a correlation of the
14,315 IRAS galaxies with the RASS~II source list.~\footnote{
In order to maintain a well defined infrared source
sample we did not include in our correlation the incomplete
sample of high flux [$f(100\mu{\rm m})>10$ Jy] IRAS galaxies. 
These objects are excluded by the multivariate selection 
technique (cf. Adorf \& Meurs 1988; Boller et al. 1992b; Sect. 2 of 
Boller et al. 1992a); however their RASS~I spatially correlated sources are 
included in our first correlation.}
For consistency we have still taken the IRAS PSC sources, but use 
whenever available the IRAS FSC position and fluxes instead of the IRAS PSC 
positions and fluxes.
Section~2 describes how the X-ray emitting
IRAS galaxies were identified through a spatial 
correlation, 
and a subsequent superposition of X-ray emission contours on
optical images. We adopted  a classification scheme
to characterize the likelihood that the infrared and X-ray emission
originate from the same object. Thereby we obtained a list of IRAS galaxies
with a very high probability of detected X-ray emission. 
Their soft X-ray spectral properties were analyzed and compared with
their infrared emission in Sect. 3.
 
The aim of this paper is to present the  basic data and to quantify
the likelihood that the X-rays we measure are associated with the 
IRAS galaxy.
In subsequent papers we shall present
results of optical follow-up observations of our galaxy sample and of
theoretical models addressing the X-ray and far-infrared emission of
galaxies in different states of nuclear activity.
 
The images for the 372 potential identifications and the tables 3 to 5 of 
the paper are available as postscript files from 
{\tt
http://www.rosat.mpe-garching.mpg.de/}
\~~bol/iras\_rassII
or can be retrieved via anonymous ftp from
{\tt 
ftp.rosat.mpe-garching.mpg.de
}
in the subdirectories 
{\tt  /outgoing/bol/iras\_rassII/images}
and
{\tt  /outgoing/bol/iras\_rassII/tables},
respectively.

\section{Identifying X-ray emitting IRAS galaxies}

Our RASS~II source catalogue only includes sources with a detection
likelihood larger than eight.
In addition, the number of source
photons in the (0.1$-$2.4 keV) energy band must be larger than six.
The latter restriction is obtained from an analysis of the distribution of
detection likelihood values versus the ratio of source to background
count rates.  Sources with five or less source photons show unusually
high source to background count ratios for their values of the
detection likelihood, compared to sources with a higher detection
likelihood. This may be due to uncertainties in the standard analysis
software analysis of weak sources. To avoid the uncertainties
introduced by the SASS detection algorithm we therefore require the
number of source photons to be above some threshold.  The RASS~II
source catalogue thereby constrained has 116,471 sources.

Through a correlation of the positions of 14,315 IRAS galaxies with 
116,471 RASS~II source positions we obtained a list of 372
possible identifications.
Because of its superior positional accuracy, when available
we used the IRAS FSC position instead of the
IRAS PSC position. Table 2 lists the corresponding FSC and PSC names
for such objects.

In Sect. 2.1 we estimate the number of chance coincidences among our
correlated IRAS and RASS~II sources, as well as the dependence of that
number on the positional uncertainty of the RASS~II source positions.

In a second step (Sect. 2.2) the candidate identifications were
individually examined through superimposing the X-ray emission
contours on optical images taken from the Palomar Digitized Sky
Survey. In these images we also show the
IRAS 3 $\sigma$ error ellipse. 
All objects were classified according to how clearly the
X-rays appear to originate from the optical counterpart of the
respective IRAS source.

\subsection{Identification through positional coincidence 
and chance coincidence rate}

\begin{figure*} 
  \begin{minipage}{6.5cm} 
       \psfig{figure=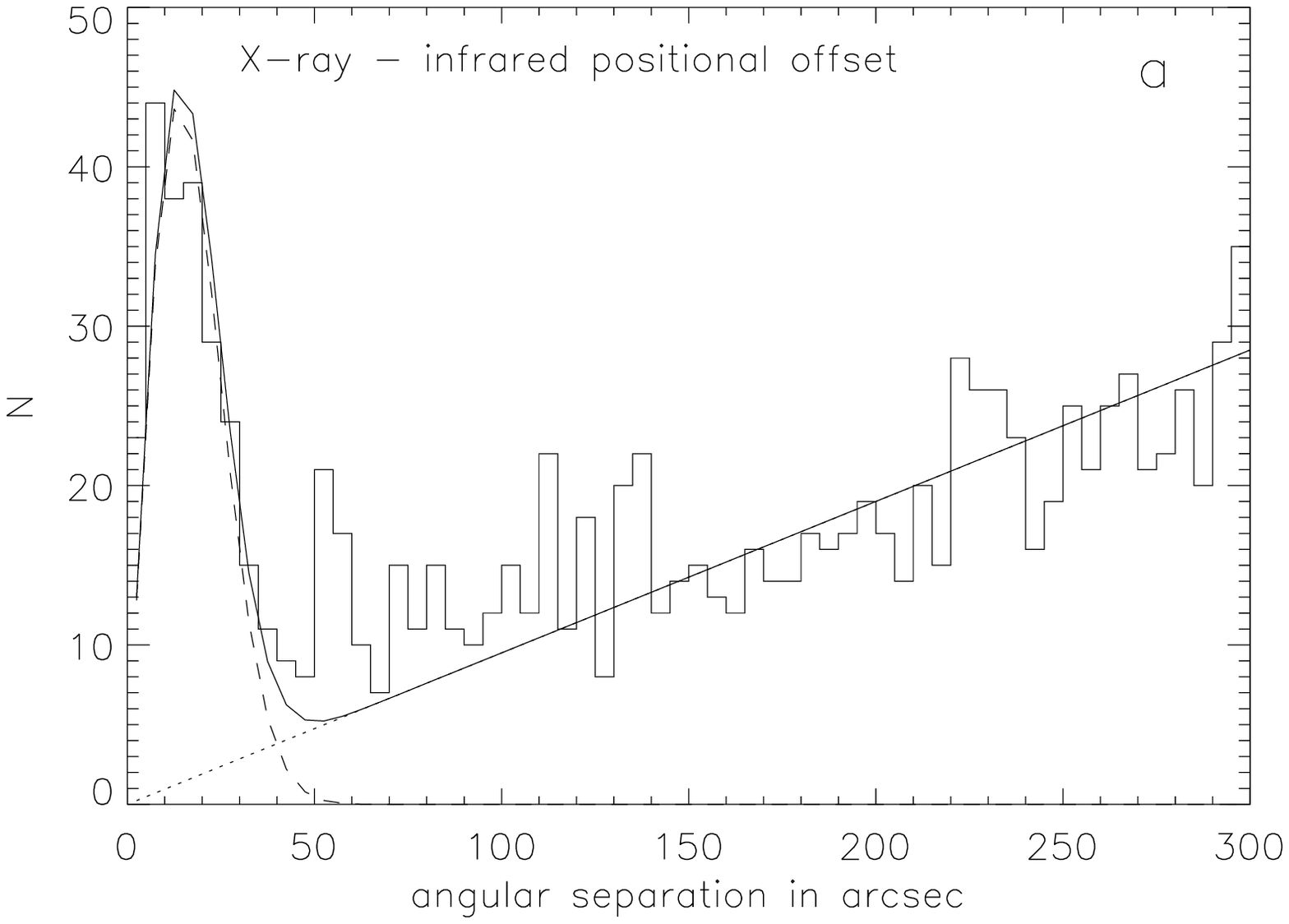,height=6.3cm,clip=}
\end{minipage}
  \begin{minipage}{6.5cm}
       \psfig{figure=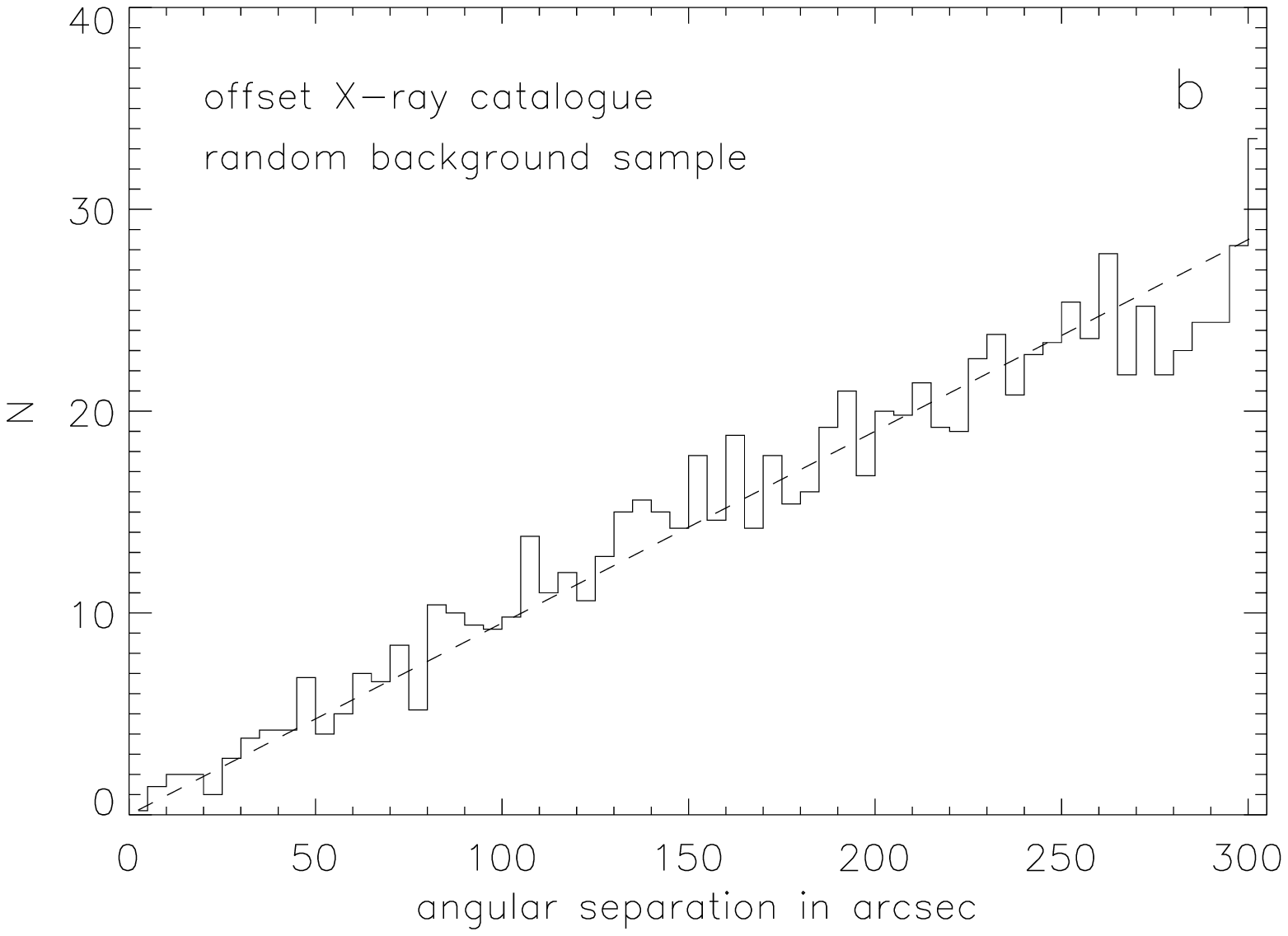,height=6.3cm,clip=}
\end{minipage}
\caption{
Distribution of separations between the 
IRAS PSC and ROSAT All Sky Survey II sources (left pannel) and
an offset X-ray catalogue (right pannel, see text for details). 
We chose a maximum angular separation of 100 arcsec to obtain
a first set of 372
candidate identifications.
We estimate between 98 and 113 chance coincidences 
among these.
After a visual examination of these objects we allow for a maximum
separation of only 30 arcsec, which reduces the number of expected
chance coincidences to $10\pm1$ pairs.
}
\end{figure*}

This section describes our first identification step, a search for
RASS~II sources within 5 arcmin around each of the 14,315 IRAS
galaxies.  We estimate the chance coincidence rate and its dependence
on the positional identification uncertainty.

\subsubsection{Chance coincidence rate from two random sets}

From two random sets of $ N_1$ = 14,315 and $ N_2$ = 116,471 
sources we expect
$$
N_{cc} = \frac{N_1\ N_2}{4\pi} \cdot \pi
\left(\rm \pi \theta\over 180\cdot 60^2\right)^2 
= 98 \left(\theta\over 100^{"}\right)^2
\eqno(1)
$$
chance pairs with an angular separation less than $\theta$,
where $\theta$ is measured in arcsec.
Within a separation of 100 arcsec we therefore expect
98 chance coincidences, which account for one quarter of
the source pairs we found.

To empirically verify this number of chance coincidences
we re-correlated the IRAS galaxy positions five times with
RASS~II positions that were offset by $\pm$30 arcmin in 
right ascension and/or declination.
The number of chance coincidences of separation less than
100 arcsec turned out to be
101 ($\Delta\alpha = +30^{'}$),
105 ($\Delta\alpha = -30^{'}$),
113 ($\Delta\delta = +30^{'}$.), 
105 ($\Delta\delta = -30^{'}$), 
102 ($\Delta\alpha = +30^{'}$, $\Delta\delta = -30^{'}$).
The slightly higher number of pairs in the test correlations may be
due to an excess of correlation below about 150 arcsec separations
that we attribute to a wider spatial correlation between diffuse
cluster X-ray emission and the infrared emission from galaxies within
the clusters (see the discussion in the following Section).


The number distribution of pairs drawn from two randomly
distributed  sets of $N_1$ and
$N_2$ objects, respectively,
should follow 
$$
\frac{dN_{cc}}{d\theta} 
= \frac{N_1 N_2}{2}\frac{\pi^2\theta}{180^2\cdot 60^4} 
= 0.0196\left(\frac{N_1}{14315}\right)\left(\frac{N_2}{116471}\right) 
\cdot \theta
\eqno(2)
$$
A least square fit to
the  average distribution up to 300 arcsec of
our five test correlations (cf. Fig. 1b) yields 
$ dN_{cc}/d\theta = \rm 0.0190\cdot  \theta$,
which corresponds to $N_1\cdot N_2 = 1.62\times 10^9$, in fair 
agreement with the 
expected value of $N_1\cdot  N_2 = 1.67\times 10^9$. 


\subsubsection{Chance coincidence rate from the RASS~II - infrared catalogue}

We searched for RASS~II counterparts within a radius of 5 arcmin
around the position of each of the 14,315 IRAS galaxies
and found 372 possible identifications.  
Fig. 1a shows the distribution of distances between the
infrared position and its nearest X-ray source.  In contrast to the
distribution of random associations displayed in Fig. 1b, the
actual correlation of the two catalogues shows a strong excess of
source pairs within about 50 arcsec.
Of
the 372 pairs within 100 arcsec the expected number of chance coincidences is 
105, with
an uncertainty estimated from the different X-ray catalogue offsets of
about $\pm$ 8.  
Thus statistically, we should have found about 267
IRAS sources with X-ray emission in their actual vicinity.

The width of the
distribution of ``real" source pair\ \  separations arises from positional
uncertainties in each catalogue, resulting in a Gaussian distribution
with some angular width $\sigma$.  
We therefore model the observed
distribution as the sum of $N_{real}$ ``real" pairs, and a background
characterized by the total number, $N_1\cdot  N_2$, of pairs in two sets of
randomly distributed objects:
$$
dN(\theta) = \left( 
\frac{N_{real}}{\sigma^{2}} e^{-\theta^2/2\sigma^2}
+ \frac{N_1\ N_2}{2} \frac{\pi^2}{180^2\cdot 60^4} \right) 
~\theta ~d\theta~ \eqno(3)
$$
A least square fit to the observed distribution
yields $\sigma=14$ arcsec  and    $N_{real}= 204$ (we have fixed the 
value of $ N_1\cdot  N_2$ to that obtained from the distribution of Fig. 1b).
Our visual inspection  (cf. Sect. 2.3 and our overlays) of all 372 pairs 
resulted in 197 secure identification of X-ray emitting
IRAS galaxies, a number in good agreement with the 204 
objects attributed to the central Gaussian of the pair distribution
function.

The number of 204 sources we attribute to the central Gaussian is smaller than
the total number of pairs to 100 arcsec minus the expected random pairs,
372 - 105 = 267.
This is probably due to 
another excess 
above the expected random distribution and outside 
the central Gaussian peak, reaching out to about 150 arcsec
(cf. Fig. 1a). 
This wider excess could most naturally be explained as coming from IRAS
galaxies embedded in clusters with diffuse X-ray emission from the
intercluster gas. 
In such cases the IRAS galaxy is not the source of the X-ray emission,
which come from the cluster gas and may peak somewhere within the cluster size
away from the IRAS galaxy.
The size of nearby clusters from which we would
expect RASS--detectable intercluster gas emission is in fact of order
2 arcmin.
In our detailed comparison of the ROSAT emission with the optical images
we indeed found such clusters
that we flagged accordingly (grade 3, cf.
Section 2.3). The X-ray emission on the line of sight to IRAS F05537$-$6653
is a representative example.

\begin{figure}
       \psfig{figure=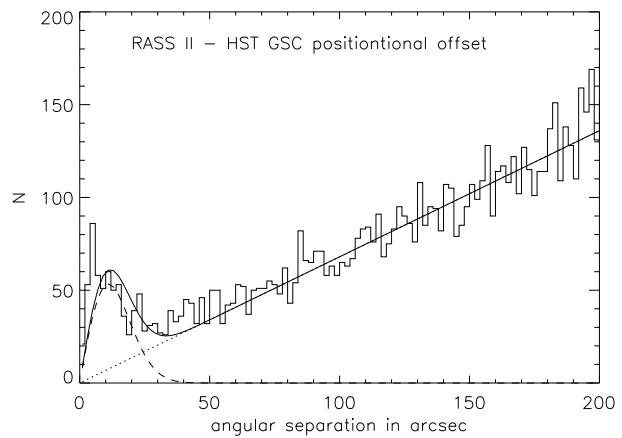,height=6.3cm,clip=}
      \caption{Distribution of separations between the 
HST GSC and RASS~II detected IRAS galaxies.
The standard deviation $\sigma$ of the Gaussian distribution is 11 arcsec.
The value of 11 arcsec can be used
to estimate the pointing accuracy for RASS~II sources.
}
\end{figure}

For our 372 candidate pairs we searched the NASA/IPAC extragalactic
database (NED) for optical identifications within a distance of 2
arcmin around the IRAS PSC positions.  Table 3 lists the X-ray and
infrared positions, and the positions of the found NED counterparts
(we list at most nine of them) with their basic optical properties.
Since NED comprises many catalogues, it may identify a given galaxy
with several different names and coordinates.  When an IRAS galaxy was
clearly identified with an optical galaxy, the more accurate optical
galaxy position is listed in Table 3 instead of the IRAS position.

We are aware that the spectral classification in NED is collected
from the literature without quantifying the likelihood of the 
classification. 
For a better and consitent classification 
we therefore initiated an optical follow-up program
the results of which will be presented in a separate paper.

\subsubsection{X-ray position accuracy}

In order to estimate the positional (aspect) errors of the X-ray
sources we selected all IRAS galaxies with optical counterparts, and
correlated the optical and RASS~II positions.  For accurate optical
positions for the IRAS galaxies we searched the {\it Hubble
Space Telescope} (HST) Guide Star Catalogue (GSC) (Lasker et al. 1990;
Russell et al. 1990; Jenkner et al. 1990). Within a search radius of 5
arcmin 16,274 counterparts were found.  Figure 2 shows the
distribution of the optical and RASS~II position offsets.  Under the
reasonable assumption that the accuracy of the HST GSC positions is
much higher than that of the RASS~II positions, we estimate the
aspect error of the X-ray positions from the Gaussian width of the
central distribution peak as $\sigma=11$ arcsec.
For a Gaussian distribution this results in 68\% (90\%) of ROSAT
sources found within 10 arcsec (18 arcsec) of the HST GSC position.
These numbers are in good agreement with the results 
(12 and 20 arcsec)
found by
Voges \& Boller (1997) by correlating the ROSAT Bright Source Catalogue
(Voges et al. 1996b)
with the TYCHO catalogue for bright stars having an error of less than 1 arcsec
(cf. their Fig. 3).


\subsubsection{Comparison with RASS~I catalogue}

The 14,315 IRAS galaxies were previously correlated with the RASS~I
source catalogue and the results were presented in Boller et al. (1992a).
The advantages of the RASS~II processing with respect to the RASS~I processing
are discussed in the introduction of this paper. 
52 objects
from the original RASS~I source catalogue are no longer included in
the present catalogue. There are three reasons why objects 
were excluded;
(i) the infrared flux
at 100 $\rm \mu m$ is greater than 10 Jansky (see the footnote in our introduction),
26 objects fulfill this criteria;
(ii) the positional offset from the new RASS~II processing between
the RASS~II position and the infrared position is larger than 100 arcsec and/or
the RASS~II detection likelihood is less than eight
(25 objects), and (iii), the number of RASS~II source photons is less than
six (one object).

Due to the merging of the original RASS~I strips into 1378 sky-fields
of 6.4 $\times$ 6.4 degree, 78 new objects at fainter X-ray fluxes are
detected in the RASS~II processing with respect to the RASS~I processing
and these objects appear now also
in our new RASS~II catalogue of IRAS galaxies.

The improved RASS~II processing and our individual 
examination of the X-ray emission structure with respect to the optically 
visible galaxy and its environment together provide a higher degree of
reliability of this catalogue compared to the previous RASS~I $-$
IRAS correlation. A third release of this catalogue might be
called for if major improvements can be made in the SASS processing.

\subsection{X-ray contours on optical images}
 
The Photon Event (PET) Files from the RASS~II processing were used to
produce X-ray images and contour plots.
To obtain an optimal spatial resolution of the X-ray images, the 
PET files were binned in 5 arcsec width bins.
The resulting  images were then smoothed with a $\sigma=19$ arcsec 
Gaussian filter, corresponding to a full-width at half maximum of 45 arcsec, 
which is the 
expected width of the point spread function in the ROSAT All-Sky Survey.
X-ray contours were computed in units of source photons per FWHM detection cell.
The background photon density $n_{bg}$ is known at each source position 
from the RASS~II processing.
To detect at least one source 
photon within a circular area of 45 arcsec diameter requires a mean
photon density of $\rm 6.29 \cdot 10^{-4}\ photons ~ arcsec^{-2} + \it
n_{bg}$.
The source plus background photon density is known from the PET 
files and the exposure map. 
The lowest contour line
was chosen to represent two source photons per FWHM detection cell.
The higher contour levels represent 3,5,9,17 
(doubling the contour value difference) and then $2^n$
(for $n\ge5$) photons per FWHM detection cell. This choice of 
contours appears to best trace  the source structure over
their dynamic range.
 
The X-ray photon flux contours were overlaid on optical images, 
and a cross marks the centroid X-ray position obtained by the
RASS~II processing. Rectangles were drawn 
to mark the positions of NED sources, and for the IRAS FSC or PSC 
sources the 3$\sigma$ error ellipse.
The likelihood grade of the association of the X-ray emission with the
IRAS galaxy (see next section) 
is printed in the lower right corner of each image.

\subsection{Classification of X-ray detection likelihood}

For a  visual inspection  we consider all 372 pairs with a 
separation of 100 arcsec as
potentially ``real" correlations subject to further analysis.

This does however not necessarily imply that the IRAS galaxy itself is
responsible for the X-ray emission. Since galaxies tend to be
clustered (cf. our discussion in Sect. 2.1.1), the IR and X-ray emission may 
arise from neighboring galaxies, or 
from 
extended intercluster gas.   
An visual inspection of all 372 sources is therefore necessary to assess the 
emission structure in relation to the optically visible galaxy and its 
environment. 
In the visual examination we finally allow for a maximum
separation of only 30 arcsec
between the X-ray and the IR position,
 which reduces the number of expected
chance coincidences to $10\pm1$ pairs.

We adopted a classification scheme with the following
six grades to characterize the 
quality and likelihood of an identification of the RASS~II
emission with an IRAS galaxy.
 
\parindent=0.0cm
\noindent
1:
The X-ray emission is spatially 
clearly coincident with an optical counterpart to the
IRAS galaxy. There is no significant surrounding
X-ray emission peak within the 5 arcmin detection cell used in the RASS~II
processing.
Objects with a classification grade 1 have the highest 
likelihood that the X-ray emission is actually originating from the 
IRAS galaxy.
 
2:
The X-ray emission is spatially coincident with 
an optical counterpart of the
IRAS galaxy. However, there is significant
surrounding X-ray emission within the RASS~II detection cell.
In this case the RASS~II count rate was corrected by subtraction of
the surrounding emission peaks, using the
PET files.
The number of source photons after subtraction of the secondary peaks
is still required to be equal or greater then six.
 
2db:
The IRAS galaxy may be a blend of two or more galaxies not
resolved in the IRAS FSC or IRAS PSC. The X-ray emission is spatially 
coincident with the infrared position. The spatial resolution of
the ROSAT All-Sky Survey does not allow a unique identification of
one of the optical counterparts with the X-ray emission.

3:
The X-ray emission shows complex and diffuse structure and cannot
be uniquely associated with the IRAS galaxy. The complex emission
may originate from intercluster gas, background or foreground 
objects. An identification with the IRAS galaxy is uncertain.

4:
There is significant X-ray emission near the IRAS position, but
it peaks at a distance larger than about 30 arcsec 
from the IRAS galaxy, which is about $2.5\sigma$ from the
X-ray peak. This appears too far, but in some cases may be 
due to pointing errors in RASS II.
 
5:
There is no spatial association of the X-ray emission with the
IRAS galaxy. The RASS~II centroid position only fortuitously
coincides with the IRAS galaxy position.

9:
Other:\\
(i) a bright foreground star close to the position of the IRAS galaxy
is the most likely source of the X-ray emission;\\
(ii) less than 2 source photons are detected at the position of the RASS~II
centroid position, i.e. no contour lines are plotted (cf. IRAS 05576$-$7655).

Only objects with a classification grade 1, 2 or 2db are 
considered secure identifications of X-ray emission with an IRAS 
galaxy.

\subsection{RASS~II sources with resolved X-ray emission}

In Table 1 we list sources considered as secure identifications
which have an source extent that exceeds the point spread function.
The likelihood of the source extent is required to be at least 10,
a value obtained from a verification process of that SASS parameter.

\begin{table}
   \caption{RASS~II sources which have a source extent 
larger than the
point spread function. Column 1 gives the source name from IRAS FSC or PSC, 
respectively. The source extent above the point spread function in
arcsec in given in column 2. The last column lists the likelihood of 
the source extent (cf. Cruddace et al. 1986). 
}
  \begin{tabular}{rrr}
    \hline
(1)        &(2)  &(3) \\
IRAS name  & extent &likelihood \\
F07387+4955 & 27 & 186 \\
F13518+6933 &13&23\\ 
F14157+2522&10&26\\
F14400+3539& 12&38\\
F16136+6550& 11&19\\
F17020+4544& 12&11\\
F18011+4246&75&12\\
F18216+6419 &23&485\\
F20240$-$5233&67&16\\
F22402+2927& 9&26\\
23566$-$0424&19&10\\
    \hline
\end{tabular}
\end{table}

\section{Soft X-ray and far-infrared properties of secure identifications}

In Table 4 we list the soft X-ray (0.1--2.4 keV) fluxes and luminosities 
and the far-infrared ($\rm 40-120 \mu m$) fluxes and luminosities
for the secure identifications.
To compute 
the soft X-ray (0.1--2.4 keV) energy flux
from the PSPC count rate we assume a simple power-law spectrum 
\begin{equation}
f_E\ dE ~\propto~ E^{-\Gamma+1}~ dE~,
\end{equation}
where $f_{E}\ dE$ is the galaxy's energy flux between photon energies
$E$ and $E+dE$. We assume a fixed photon spectral index $\Gamma=2.3$,
which is the typical value found for extragalactic objects with ROSAT
(cf. Hasinger et al.  1991, Walter \& Fink 1993), and
an absorbing column density of hydrogen fixed at the 
Stark et al. (1992) Galactic value $N_{\rm H gal}$ along the line of sight.
In other words, the normalization of the spectrum from which the energy flux 
is derived is chosen such that a power law photon spectrum with 
$\Gamma=2.3$, absorbed by a column $N_{\rm H gal}$, produces the
observed count rate. These fluxes are referred to in the following as
$f_{\rm X1}$.

For 42 objects with more than 100 detected source photons we were able
to improve the flux estimates by fitting power-law spectra with 
free spectral index $\Gamma$, and free absorption column density, $N_{\rm
H}$; latter was however required to be larger or equal to the Galactic
value at the respective position, $N_{\rm H gal}$. 
Under the assumption that the intrinsic spectrum is a power law,
we can thereby account for absorption of X-rays 
inside the respective galaxy. 
The flux derived this
way, $f_{\rm X2}$, should usually be higher than that derived with fixed
$\Gamma$ and $N_{\rm H}$, since 
it is derived from a larger absorbing column.
However, in some cases $f_{\rm X2}<f_{\rm X1}$. This occurs when our best
fit $\Gamma<2.3$, which results in a lower energy flux for a 
given count rate, and when at the same time 
$N_{\rm H}$ is not much larger than the Galactic value.
It may also be 
due to differences between the count rates
determined in the standard analysis and the count rates derived  from
the Photon Event files using the EXSAS package (Zimmermann et al.
1994).
The integrated energy fluxes and other 
spectral parameters for the 42 X-ray brightest galaxies we were able 
to fit this way  are listed in Table 4.  
In Fig. 3 we compare the fluxes obtained from a free fit, $f_{X2}$, with
those of a fixed spectral shape fit, $f_{\rm X1}$. As expected, most objects
have $f_{\rm X2}>f_{\rm X1}$. 

\begin{figure}
       \psfig{figure=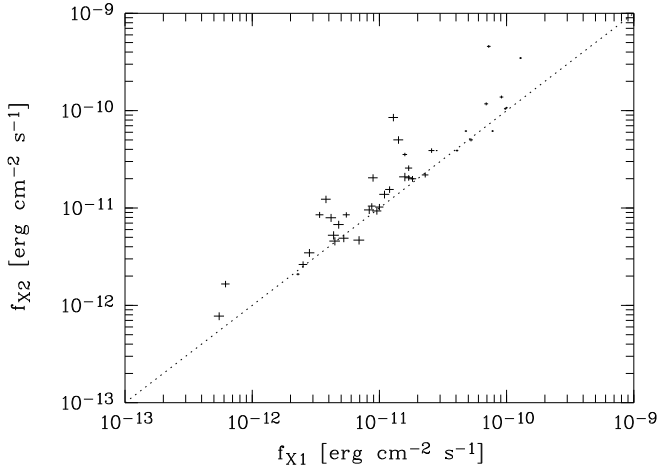,height=6.3cm,clip=}
      \caption{Soft X-ray flux $f_{\rm X1}$ derived from a power-law model
with fixed spectral index $\Gamma=2.3$ and corrected only 
for Galactic absorption, plotted against
the flux $f_{\rm X2}$ computed from a best fit power-law spectrum with free 
spectral index and absorbing column $N_{\rm H}\ge N_{\rm H gal}$.
}
\end{figure}
\begin{figure}
       \psfig{figure=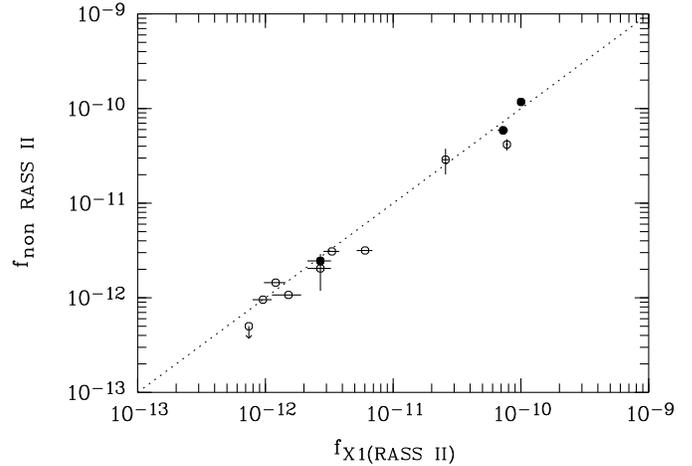,height=6.3cm,clip=}
      \caption{
Comparison of the X-ray fluxes $f_{\rm X1}$ obtained in this paper
with X-ray fluxes obtained by other authors. We searched the literature
and found 11 sources out of the 197 secure identifications listed in 
Table 4 with published soft X-ray fluxes.
The open symbols mark detections by the Einstein
satellite and the filled symbols were obtained within ROSAT pointed 
observations.
The fluxes obtained from different measurements are in good agreement
and the differences 
may be due to either
intrinsic variability
of the sources (cf. Voges \& Boller 1997 for an statistical analysis
of the variability of ROSAT sources) or slight differences in the
spectral modeling used to convert count rates to fluxes. 
}
\end{figure}

We are aware that in objects where detailed spectral modeling was
performed, it often though not always appears that the spectrum can be fit
with a hard power-law plus a soft excess component. Even if the spectrum
is well represented by a power-law, the spectral index may assume a wide
range of values (e.g. Boller et al. 1996) find a correlation
between the soft X-ray photon index and the FWHM of the $\rm
H_{\beta}$ line). 
In these cases the power-law approximation may result to uncertain
fluxes and luminosities.

To illustrate the robustness of our flux determination we have
searched the literature for detections by other 
X-ray satellites or other authors on RASS-detected IRAS galaxies from
this paper.
From the secure identification listed in our Table 4 the following sources
have published X-ray fluxes obtained with the 
Einstein satellite (Fabbiano et al. 1992):
IRAS F02321-0900, 
F03207-3723 (as well as  ROSAT RASS~I observations (Brinkmann et al. 1994)),
F03372-1850,
F04150-5554,
F11034+7250,
F11210-0823,
F12125+3328,
F13277+5840,
F14157+2522;
and within ROSAT pointed observations (Brinkmann et al. 1994)
F04305+0514 and 
F12265+0219.
Figure 4 shows a good agreement between different flux measurements.
We are  aware that source variability might also contribute
to the scatter in Fig. 4, since the majority of ROSAT sources show
variability (cf. Voges \& Boller 1997 for a statistical analysis
of the variability of ROSAT sources).

The total far-infrared (40-120$\mum$)
fluxes, $f_{\rm FIR}$, were computed following Helou 
(1985) from the IRAS 60 $\mu$m   and 100 $\mu$m 
band fluxes: 
$$
f_{\rm FIR} =1.26\times 10^{-11}  (2.58 f_{60}+f_{100})  
\rm \ erg\ cm^{-2}\ s^{-1},\eqno(2)
$$
where $f_{60}$ and $f_{100}$ are given in Jansky.
The soft X-ray and far-infrared fluxes were converted to luminosities using
equation (7) of Schmidt \& Green (1986):
\addtocounter{equation}{1}
\begin{equation}
L(E_1,E_2) = 4 \pi (c/H_0)^2\ C(z)\ A^2(z)\ f(E_1,E_2),
\end{equation}
where a power-law spectrum is assumed in the energy range $(E_1,E_2)$,
so that the redshift-dependent
 functions $C(z)$  and $A(z)$ are then given by:
\begin{eqnarray}
C(z) &=& ( 1 + z  )^{\Gamma - 2},
\\ 
A(z) &=& 2 \left[(1+z) - (1+z)^{1/2} \right].
\end{eqnarray}
For the photon index in the far-infrared we assumed
$\Gamma= 1.5$. 
A Hubble constant $H_0 = 50~ \rm km\ s^{-1}\ Mpc^{-1}$ and
cosmological deceleration parameter of $ q_0 = \rm 0.5$ were adopted.

\begin{figure*}
\begin{center}
       \psfig{figure=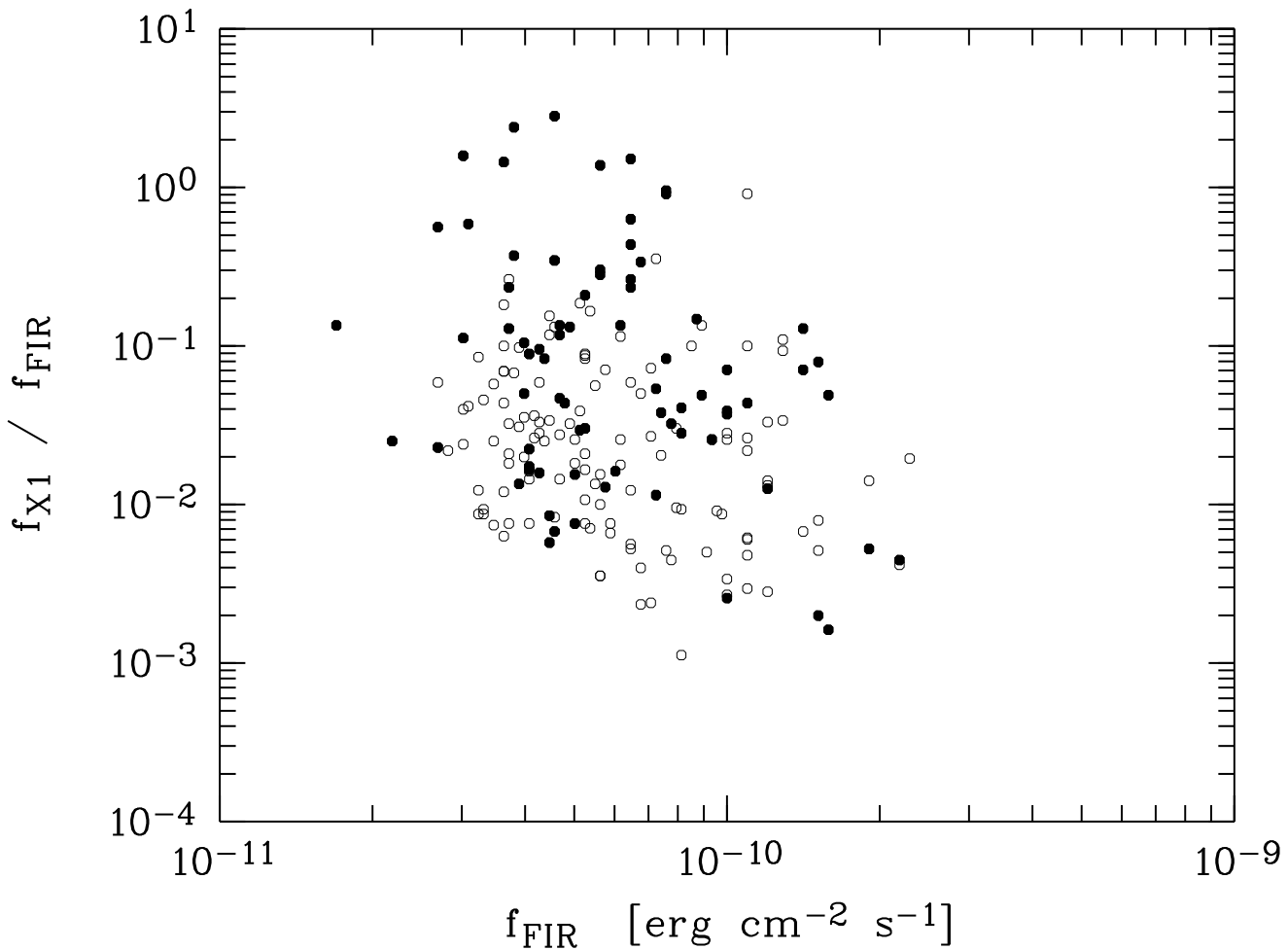,height=13.5cm,width=18cm,clip=}
      \caption{Soft (0.1--2.4 keV) 
X-ray to far-infrared (40--120$\mum$) flux ratio versus far-infrared flux for
ROSAT All-Sky Survey detected IRAS galaxies. Only objects with
confident correlations, i.e., grade 1 and 2, are shown. Galaxies identified
as Seyfert in NED are marked by filled symbols.
The X-ray flux was obtained from a simple power-law model with a 
photon index $\Gamma=2.3$ and an absorbing column density of hydrogen
fixed to the Galactic value along the line of sight (cf. Sect. 3).
}
\end{center}
\end{figure*}

In Fig. 5 we plot the ratio between
the far-infrared and soft X-ray flux, $f_{\rm X1}/f_{\rm FIR}$ 
against the far-infrared flux for the
objects in Table 4.
Galaxies identified as Seyfert in NED are marked as filled symbols.
The ratio $f_{\rm X1}/f_{\rm FIR}$ ranges over 4 orders of magnitude and 
IRAS galaxies
identified as Seyfert in the NED tend to be located at the high
end of this distribution.

\section{Summary}

(1)
We have correlated the  116,417 X-ray sources detected in the RASS~II processing
with a sample of 14,315 galaxies selected from the IRAS PSC. Our results 
build upon a previous correlation of these IRAS galaxies with the RASS~I
processing. 

(2)
372 IRAS galaxies show X-ray emission within a distance of 100 arcsec
from the infrared position.

(3)
X-ray contour plots overlaid on Palomar Digitized Sky Survey images were
inspected to quantify the likelihood that the X-rays originate from the
IRAS galaxy. These images are shown for each of the 372 potential
identifications.

(4)
For 197 objects the soft X-ray emission is very likely associated with the
IRAS galaxy. For these objects the soft X-ray fluxes and luminosities
are computed and compared with their far-infrared emission. 
For 42 objects with more than 100 detected source photons we were able
to improve the flux estimates by fitting power-law spectra with 
free spectral index  and free absorption column density.

(5)
The ratio between the soft X-ray and the far-infrared flux covers about
4 order of magnitudes and can reach values above unity.

The 197 IRAS galaxies with secure X-ray emission are subject of
an optical follow-up observation program. The results will be presented
in subsequent papers.

\acknowledgements
We thank J. Tr\"umper, W. Pietsch and U. Zimmermann for helpful discussions.
We thank the referee, Dr. M. Lehnert for his suggestions to
further improve the paper.
The  ROSAT project is supported by the Bundesministerium
f\"ur Bildung, Wissenschaft, Forschung und Technologie (BMBF/DARA) 
and the
Max-Planck-Society.
FB acknowledges support by the Deutsche Forschungsgemeinschaft. 
 
{}
 
\vfill \eject

\appendix
\section{Notes on individual objects}

IRAS  F01268$-$5436\\
Two galaxies are located within the IRAS FSC 3$\sigma$ error ellipse.
The major X-ray emission is probably associated with the galaxy
centered south-east of the IRAS centroid position.
\vskip 0.2cm

IRAS  F02562+0610\\
Several optical sources in or near IRAS 3$\sigma$ ellipse, and
several optical sources near X-ray emission, thus no
unique identification possible.
\vskip 0.2cm

IRAS  F04392$-$2713\\
X-ray centroid and NED position for IRAS  F04392$-$2713 agree well,
however, the IRAS FSC error ellipse is centered on a nearby galaxy about 1 arcmin
west of the NED position, which is probably the source of the infrared emission.
\vskip 0.2cm


IRAS  F05136$-$0012\\
Relatively large offset of about 50 arcsec between the X-ray centroid
and the infrared position.
A PSPC pointed observation on F05136$-$0012 indicates that the X-ray
centroid position and the optical position coincide. 
Therefore the offset is probably due to an unusually
large aspect error of the survey data.
\vskip 0.2cm


IRAS  F05340$-$5804\\
The NED IRAS position is taken from the PSC and is inaccurate.
The FSC position is associated with an optical galaxy, but this
galaxy is not the source of the
X-ray emission.
\vskip 0.2cm

IRAS  F06059$-$4928\\
The infrared emission is not clearly associated with any particular
optical galaxy of the apparent cluster.
NED identifies  F06059$-$4928 
with one of the galaxies but the X-ray emission
is clearly not associated with that object.
\vskip 0.2cm

IRAS  F06068$-$2705\\
Condon et al. (1997) find that the IRAS and ROSAT sources are not
identical. Most X-ray emission is probably related to the
northeastern member of the galaxy pair near the X-ray centroid position
which is a Seyfert 1.9 galaxy.
\vskip 0.2cm

IRAS   06269$-$0543\\
Multiple optical counterparts within IR 3$\sigma$ ellipse.
\vskip 0.2cm

IRAS 08140+7052\\
Irregular galaxy Ho II with about 7 arcmin diameter, 
multi-component X-ray emission.
Classification grade 1 is given 
since all the emission originates inside the galaxy
despite the large
offset between the IRAS FSC and the X-ray centroid
positions.
The infrared luminosity may be larger than given by
FSC/PSC.

\vskip 0.2cm

IRAS  F10214+0644\\
Condon et al. (1997) identify the galaxy at
optical position 
$\rm \alpha = 10^h23^m59.9^s, \delta = +06^{o}29^{'}00^{''}$
as the IRAS and VLA source. The ROSAT source
is probably associated with the radio-quiet galaxy CGCG 037$-$022.
\vskip 0.2cm

IRAS   11598$-$0112\\
Two optical counterparts within IRAS PSC
3$\sigma$ error ellipse. The X-ray emission is not spatially resolved.
\vskip 0.2cm

IRAS  F12134+5459\\
The IRAS source may be a blend of MCG +09$-$20$-$133 and
MCG +09$-$20$-$134. 
From the RASS~II contour lines we cannot resolve which galaxy
is actually the source of the X-ray emission.
\vskip 0.2cm

IRAS  F13429+6652\\
A nearby optical counterpart to IRAS galaxy could contribute
to X-ray emission. The X-ray emission from the two components
is not spatially resolved.
\vskip 0.2cm

IRAS  14072$-$5205\\
The bright optical counterpart near the 3$\sigma$ IRAS PSC
error ellipse may be a star. 
The IRAS galaxy is not listed in the FSC.
\vskip 0.2cm

IRAS  F15195+5050\\
Condon et al. (1997) identify the Seyfert 1 
CGCG 274$-$040 NED02 
at optical
position $\rm \alpha = 15^h21^m8.9^s, \delta = +50^{o}40^{'}07^{''}$
as the probable source  of the X-ray emission.
The starburst galaxy CGCG 274$-$040 NED01 is the
primary identification of the VLA source.
The X-ray emission is clearly identified with a neighboring optical
galaxy. Both the VLA and ROSAT galaxies are equally distant to
the IRAS source, thus it remains unclear which is the IRAS source.
\vskip 0.2cm


 IRAS  F17023$-$0128\\
Following Condon et al. (1997) and references for object note on 
IRAS  F17023$-$0128,
the Seyfert 1 galaxy UGC 10683 NOTES02 is the most likely 
IRAS source. This source is also the main X-ray emitter.
\vskip 0.2cm

IRAS  F18011+4246\\
The peak X-ray emission appears  associated with 
CGCG 227$-$016 NED02, however the IRAS galaxy identified by NED
as CGCG 227$-$016 NED01 is within the X-ray positional error  also consistent
with the peak X-ray emission. The X-ray emission is spatially
unresolved from both components.
\vskip 0.2cm

IRAS   18396-3535 \\
The bright optical counterpart within the 3$\sigma$ IRAS PSC
error ellipse could be a star. 
The IRAS galaxy is not listed in the FSC.
\vskip 0.2cm

IRAS   19211$-$2855\\
Multiple optical counterparts within the 3$\sigma$ IRAS PSC error ellipse.
The X-ray emission could arise from a star within the IRAS error ellipse.
The IRAS galaxy is not listed in the FSC.
\vskip 0.2cm

IRAS  F19462$-$5843\\
The optical galaxy ESO 142$-$ G047 associated with the IRAS source seems
not to be related to the X-ray centroid position. There are faint
optical counterparts located at the X-ray peak emission. The ROSAT
survey data do not allow a unique identification of the IRAS source
as X-ray emitter.
\vskip 0.2cm

IRAS  F20547$-$4849\\
The optical galaxy NGC 6987 has a relatively large offset to the X-ray centroid position.
The ROSAT
survey data do not allow a unique identification of the IRAS source
as X-ray emitter. 
\vskip 0.2cm

IRAS   22146$-$5955\\
The X-ray centroid position is outside the 3$\sigma$ IRAS PSC error ellipse.
There seems no unique association of the strong X-ray emitter with one optical
counterpart.
\vskip 0.2cm

IRAS  F23251+2318\\
The X-ray emission may arise from the faint optical counterpart near the X-ray
peak emission. 


\section{Identical objects with different FSC and PSC names}

Where possible we use the IRAS FSC position and band fluxes instead
of the IRAS PSC entries. Due to an improved position determination in the
IRAS FSC the IRAS names for identical objects may differ between the
PSC and the FSC.
Below we list PSC and FSC names for identical objects.

\begin{table*}
   \caption{FSC names and PSC names for identical objects
}
  \begin{tabular}{rr|rr|rr|rr}
    \hline
PSC name   &FSC name  & PSC name    &FSC name     &PSC name   &FSC name  & PSC name    &FSC name  \\
00127+2817 &00128+2817& 00360$-$2432&00361$-$2432 &00488+2907 &00489+2908& 00540$-$0133&00541$-$0133 \\
01134+3046 &   01134+3045 &01464+1249&    01464+1248& 02025+0941&    02025+0940& 02093+3714 &   02092+3714\\
02223$-$1922&  02223$-$1921& 02537$-$1641&  02536$-$1641& 03208$-$3723&  03207$-$3723& 03229$-$0618&  03229$-$0619\\
03398$-$2124&   03398$-$2123& 03543$-$7216&  03544$-$7216& 04105$-$6811&  04105$-$6810& 04339$-$1028&  04340$-$1028 \\
04384$-$4848&  04384$-$4849& 04470$-$6227&  04469$-$6227& 04503+0114 &   04504+0114&  04576+0912 &   04575+0912 \\
05264$-$3936&  05263$-$3937& 05291$-$2608&  05290$-$2608&   05335$-$7359&  05334$-$7359 & 05399$-$8345&  05400$-$8345\\ 
05339$-$5804&  05340$-$5804& 05576$-$7655&  05577$-$7655&  05581$-$5907&  05580$-$5907& 06280+6342 &   06279+6342 \\
06295+5743  &  06296+5743&  07388+4955&    07387+4955&  07451+5543 &   07452+5543 & 08080$-$6109&  08081$-$6109\\
08066$-$1905&  08068$-$1906& 08082+7900&    08080+7900&  08162+2717&    08162+2716 & 09162+2628&    09161+2628\\
09241+5735&    09242+5735 & 09571+8435  &  09572+8435& 10213+0644&    10214+0644&  10291+6517&    10290+6517\\
11029+3130 &   11028+3130 & 11033+7250  &  11034+7250& 11058+7159&    11060+7158& 11161+6020 &   11162+6020\\
11395+1033  &  11396+1033 & 11419+2022  &  11419+2023& 12055+6527&    12056+6527& 12175+2933&    12176+2933 \\
12409+7823  &  12409+7824& 12446$-$4058&  12446$-$4057& 12551+2040&    12552+2039&  12561+2752&    12561+2751 \\
12566+3507  &  12566+3506& 12584+2803 &   12585+2803&  13003+8017 &   13004+8017&  13150+2051 &   13150+2052 \\
13177$-$2021&  13176$-$2020& 13428+6652&    13429+6652&  13446+1121&    13445+1121&  13467+4014&    13468+4013 \\
13503+6104  &  13504+6104& 13519+6933 &   13518+6933&  13510+0442 &   13509+0442&  13578+0516 &   13577+0517 \\
14156+2522  &  14157+2522& 14457+2308 &   14457+2309&  15480+6822 &   15480+6821&  15519+1444 &   15519+1445 \\
16040+1818  &  16042+1824& 17166$-$7536&  17167$-$7536&  17520+3250&    17520+3249&  17550+6520&    17549+6520 \\
17552+6209  &  17551+6209& 17511$-$6542&  17512$-$6542& 18011+4247 &   18011+4246&  18130+5703 &   18129+5703 \\
18216+6418  &  18216+6419& 18396$-$6225&  18402-6224 & 19073$-$5257&  19074$-$5257&  19290+5830&    19289+5830 \\
19463$-$5843&  19462$-$5843& 19519$-$8141&  19518$-$8142& 20051$-$1117&  20050$-$1117&  20069+5929 &   20068+5929\\ 
20044$-$6114&  20045$-$6114&  20448+2515 &   20448+2514&  20546$-$4849&  20547$-$4849 & 21236$-$6013&  21235$-$6013\\
21467$-$5530&  21468$-$5530&   22403+2927 &   22402+2927 & 22453$-$1744&  22454$-$1744& 22482$-$7027&   22481$-$7028\\
22537$-$6511&  22537$-$6512& 23016$-$5144&  23017$-$5144 & 23229+2835  &  23229+2834& 23252+2318  &  23251+2318 \\
23410+0228   & 23411+0228 & $-$ & $-$ & $-$ & $-$ & $-$ & $-$ \\ 
    \hline
\end{tabular}
\end{table*}

\end{document}